\documentstyle[twoside,fleqn,espcrc2,epsf]{article}

\newcommand{\AmS}{{\protect\the\textfont2
  A\kern-.1667em\lower.5ex\hbox{M}\kern-.125emS}}

\title{Determination of the
Crumpling Fractal Dimension via $k$-space MCRG.}

\author{{D. Espriu
\thanks{Research supported in part by EU CHRX-CT93-0343 and CICYT AEN95-0590
grants.} and
A. Travesset\thanks{Supported by a FPI fellowship from Generalitat de
Catalunya.}} \address{D.E.C.M. and I.F.A.E., Universitat de Barcelona,
 Diagonal 647,
 E-08028 Barcelona.}}

\begin{document}

\begin{abstract}
Motivated by the successful application of MCRG in momentum space to
$\lambda \phi_3^4$, we determine
the critical exponents at the crumpling transition in  fixed triangulation
surfaces. The results are still
tentative, but suggest that
$-1.0 \ge \eta \ge -1.3$, pointing at a value for the fractal Hausdorff
dimension
at the crumpling transition fixed point somewhere between 3 and 4.
\end{abstract}

\maketitle

\section{INTRODUCTION}

We have to go back to the late eighties\cite{Nelson,BEW} to find the first
evidence for
the existence of a phase transition separating the crumpled
and smooth regimes in mathematical models of surfaces
defined on a mesh with fixed connectivity. However,
the actual values of the
critical exponents associated to this phase transition remain largely
unknown. Their determination is crucial to identify which type of
conformal theory one gets in the vicinity of the critical point.

Two recent numerical simulations report the following
values for the critical exponents $\alpha$ and $\nu$. In
\cite{Wheater} a direct measuremt of the correlation length gives
$\nu=0.71\pm 0.05$ and $\alpha=0.58\pm 0.1$ (the latter is obtained via
scaling relations).
A direct measurement of the finite size scaling of the specific heat
gives $\nu=0.73\pm 0.06$, in agreement with the previous values.
In \cite{Peterson} finite size scaling gives $\omega=\alpha/\nu=1.11
\pm 0.1 $, that
is $\alpha=0.71\pm 0.05$, $\nu=0.64\pm 0.02$.
However these values
do not agree with a direct fit to the specific heat
results on large lattices (up to $128^2$), which tend to give much lower
values for $\alpha$ \cite{WS,BET}.

Virtually nothing is known of the other critical exponents
$\eta$ and $\gamma$,
which cannot be
deduced from  $\nu$ and $\alpha$ alone using scaling relations.
The most interesting one
is perhaps $\eta$,
the anomalous dimension of the field $x_i$ that describes
the position of the surface.
The fractal dimension $d_H$ is related to $\eta$
via the relation $d_H=-4/\eta$. Sometimes in random surface theory
the field anomalous dimension is introduced
through the tangent-tangent two-point function.
If we denote the corresponding anomalous dimension by $\bar{\eta}$
it follows that $\eta=\bar{\eta}-2$ and $d_H=4/(2-\bar{\eta})$.

In this note we present our preliminary results concerning the above
critical exponents obtained via the Monte Carlo renormalization
group. The action we shall study is

\begin{equation}
S=\sum_{i,j}(x_i-x_j)^2 +
\kappa
\sum_{I,J}(1-n_I n_J)
\end{equation}
with the sums extending over all pair of neighbouring sites ($i,j$)
and triangles ($I,J$). $n_K$ is the normal vector to the
$K$-th triangle. Sites and triangles live on a two dimensional
triangulated surface of fixed connectivity embedded in
${\bf R}^3$.

\section{THE METHOD}

Recently we proposed a method of implementing the renormalization
group ideas directly in momentum space\cite{ET}. It works
very well in
$\lambda \phi^4_3$ yielding critical exponents
at the Wilson fixed point with a precision
comparable,
if not better, to any other technique we know of.

Extremely long autocorrelation times difficult
the application of local
Monte Carlo algorithms to the problem at hand. For
a $128^2$ system autocorrelation times can be as large as $10^6$
sweeps\cite{Falcioni}.To cure these difficulties a Fourier accelerated
Langevin algorithm was proposed in \cite{Batrouni} and
used in random surfaces in e.g. \cite{WS}.

Fourier acceleration works best if the Langevin equation is
transformed to $k$-space.
It is therefore
natural to attempt a blocking procedure in momentum space directly. The
renormalization group transformation that we use is probably the
simplest one; we just discard half the Fourier modes in each direction.
The interested reader may wish to consult \cite{ET} for
a more detailed discussion of the method,
here we shall dwell on the details
specific to random surfaces.

In Fourier accelerated Langevin algorithms one uses the freedom
to select the two-point function for the gaussian stochastic
noise that best serves the purpose of updating the different
Fourier modes efficiently.

\begin{equation}
\langle \eta(p,\tau_1) \eta(p,\tau_2)\rangle
= \delta_{\tau_1,\tau_2} \delta_{p+k,0} \epsilon(p)
\end{equation}
For a free field theory the choice $\epsilon(p)=
(p^2+m^2)^{-1}$, where $m$ is the inverse correlation length
of the system, works best.
This also seems to be a good choice
for $\lambda \phi^4_3$.
 It has been suggested\cite{Wheater,WS} that an effective
action that describes many features of random surfaces is
$S_{eff}= -x\Delta x + \lambda x \Delta^2 x $ and thus
it would seem that

\begin{equation}
\epsilon(p)={{p^2(p^2+m^2)\vert_{max}}\over{ p^2(p^2+m^2)}}
\end{equation}
with $m$ being again the inverse correlation length, would be
a sensible choice for $\epsilon(p)$
(actually, of course, the lattice transcription of the above).
This expectation
is not borne
out by actual simulations, however. It turns out to be nearly
impossible to get good convergence of the method if
one uses the actual inverse correlation length.
We have found more convenient to
select a  relatively large value for the mass, such as
$m=1$ which seems to provide a good balance between the need to update
efficiently the slow modes and the need to keep their
stochastic excursions in check. Most of the results we present
below are obtained with this value for $m$.

A potentially more serious drawback of the method is the following.
When we update the position of the variable on the $i$-th
site we have to compute

\begin{equation}
n_I{{\partial n_J}\over {\partial x}}\propto (A_J)^{-1}
\end{equation}
If it just happens that $A_J$
is small, the field variable will receive
a large kick. Because the resulting configuration
is far from the classical trajectory, it will be promptly
brought back to order by the Langevin algorithm itself, but
such an unphysical large fluctuation spoils the statistical
samples. It should be emphasized that this is a problem
associated to the finiteness of the Langevin step
and not an intrinsic difficulty of our system. One can indeed
easily check that reducing the Langevin step reduces these
off-equilibrium excursions, but there is a limit to the
reduction of the time step as this also increases the
autocorrelation times. To alleviate this problem we
modify the action in the following form

\begin{equation}
n_I n_J\to n_I n_J \exp(-s \Delta t (A_I^{-1}
+A_J^{-1}))
\end{equation}
Table 1 illustrates the dependence of the results on the
parameter $s$ for fixed values of $\Delta t$. Some observables
can be calculated exactly and so we know when the method
reproduces the right results. We have also analyzed the
volume dependence of $s$.

\begin{table}
\setlength{\tabcolsep}{1.5pc}
\catcode`?=\active \def?{\kern\digitwidth}
\caption{Dependence on the $s$ parameter.}
\label{tab:spar}
\begin{tabular}{rrr} \hline
\multicolumn{1}{c}{\it s} & \multicolumn{1}{c}{$\langle A\rangle$}
& \multicolumn{1}{r}{
$\langle A^2\rangle -\langle A\rangle^2$} \\ \hline
\multicolumn{3}{c}{$16^2, 3\times 10^5 \ {\rm sweeps},
\Delta t=6\times 10^
{-5} $ } \\ \hline
  1     &  379.6 &  426.6 \\
  2     &  380.9 &  429.7 \\
 10     &  380.5 &  398.4 \\
Exact  &  382.5 &  382.5 \\ \hline
\multicolumn{3}{c}{$32^2, 5\times 10^5 \
{\rm sweeps}, \Delta t=8\times 10^{-5} $}
\\ \hline
  0.1   & 1542.1 & 1788.1 \\
  0.3   & 1539.8 & 1577.6 \\
  1     & 1539.2 & 1539.9 \\
  5     & 1538.4 & 1512.8 \\
 10     & 1538.4 & 1498.3 \\
100     & 1536.2 & 1485.4 \\
Exact   & 1534.5 & 1534.5 \\ \hline
\end{tabular}
\end{table}

We will discuss next the choice
of $\Delta t$ and its influence on the
autocorrelation time $\tau$. The longest autocorrelation
time is always that of the gyration radius, but we present results
for a fairly characteristic one,
 namely that of the extrinsic
curvatute $S_{EC}=
\sum_{I,J}(1-n_I n_J)$. Figure 1 shows the dependence
of $\tau$ on $\Delta t$. (The dependence on $s$ for
two values of $m$ is also illustrated --- there is
virtually none.). As expected $\tau\sim (\Delta t)^{-1}$.
In view of what we know about the extremely long autocorrelation
times in local Monte Carlo algorithms the present autocorrelation
times are truly impressive.

\begin{figure}
\vspace*{-4mm} \hspace*{+0.5cm}
\begin{center}
\epsfxsize = 0.40\textwidth
\leavevmode\epsffile{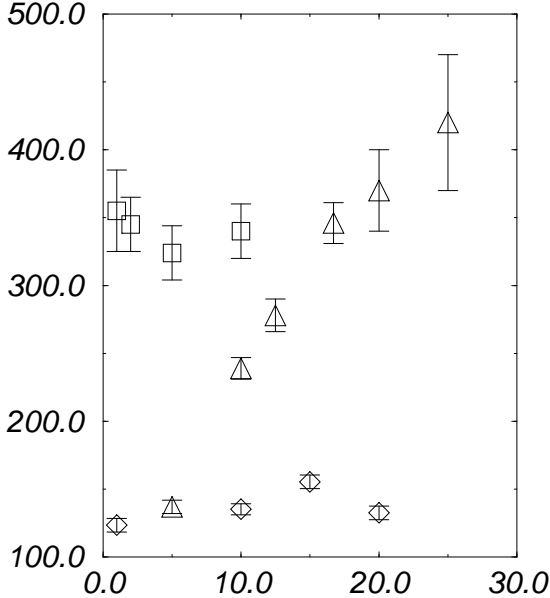}
\end{center}
\vspace{-1cm}
\caption{Autocorrelation times for $S_{EC}$ at
$\kappa=0.80$, $s=10$, $m=1$
as a function of $(\Delta t)^{-1}$ in units of $10^3$
(triangles). The squares show the dependence on $s$ of the
autocorrelation time for $\Delta t= 6\times 10^{-5}$,
 $\kappa=0.80$ and
$m=1$, and  correspond (left to right) to $s=1$,
2, 5  and 10. The diamonds
show the dependence on $s$ of the autocorrelation
 time for $\Delta t=8\times
10^{-5}$, $\kappa=0.80$ and $m=0.125$, and
correspond (left to right) to
$s=1$, 10, 15, 20.
In all cases the size is $16^2$ and the
statistics $5\times 10^5$ sweeps.}
\label{fig:1}
\end{figure}

One point that practitioners of the Langevin algorithm always
have to bear in mind is that the method is exact only
in the $\Delta t\to 0$ limit. Indeed the equilibrium
distribution is not the one corresponding to $S$ but rather
to

\begin{equation}
\bar{S}= S + {1\over 4}\sum_{i,j} \epsilon_{i,j}
{{\delta^2 S}\over{\delta x_i\delta x_j}}+ ...
\end{equation}
Even after tunning $s$, some systematic errors
remain.
Fortunately they can be reduced by working with a small enough value of
$\Delta t$. Yet $\Delta t$ has to be large
enough so that we get reasonably short autocorrelation
times and manage to sample configuration
space. After a careful analysis we have settled for
$\Delta t= 8\times 10^{-5}$ as the optimal value.
The renormalization of the parameters in $\bar{S}$ is
negligible with this value for $\Delta t$.
A detailed analysis will be provided elsewhere.
(Alternatively, one could work with a second order
Langevin algorithm, where errors are at least of
${\cal O}((\Delta t)^2)$. We have tried this, but found
no real gain.)

\subsection{The Fixed-point Action}

Now we proceed to discuss the essence of the method.
To determine the critical exponent $\nu$ we need to
determine (see e.g. \cite{ET}) the largest eigenvalue
of the matrix $T$ that linearizes the renormalization
group transformation in the vicinity of the fixed point
associated to the crumpling transition. In practice
truncation in the number of operators is required.
We have considered a total of nine operators. In continuum
notation they are
$x\Delta x$,
$x\Delta^2 x$ and
$(\partial x \partial x)^2$, and those
constructed with

\begin{equation}
{\rm Tr} K = \partial n \partial x
\end{equation}
\begin{equation}
{\rm Tr} K^2 =\partial n \partial n  =S_{EC}
\end{equation}
\begin{equation}
{\rm Tr} K^3 = (\partial_a n \partial_b n)(\partial_b n
\partial_a x)
\end{equation}
\begin{equation}
{\rm Tr K^4} = (\partial_a n \partial_b n)(\partial_a n \partial_b n)
\end{equation}
\begin{equation}
{\rm Tr}(\nabla K)^2 = \Delta n \Delta n
\end{equation}
\begin{equation}
{\rm Tr} \nabla K = \partial_a x \partial_a\partial_b n
\end{equation}
 and having dimension $d\le 2$.

\begin{table}
\setlength{\tabcolsep}{1.pc}
\catcode`?=\active \def?{\kern\digitwidth}
\caption{Comparison between `big' and `small' lattices.
$10^6$  sweeps, $\Delta t= 8\times 10^{-5}$}
\label{tab:compar}
\begin{tabular}{rrr} \hline
\multicolumn{1}{r}{per site} &\multicolumn{1}{c}{big}
&\multicolumn{1}{c}{small} \\ \hline
\multicolumn{1}{r}{$\lambda \phi^4_3$} &\multicolumn{1}{r}{$ m^2=-3.32$}
& \multicolumn{1}{c}{$\lambda=6.0$} \\ \hline
$\langle\phi^2\rangle$ & 0.272 &  0.267  \\
$\langle\phi^4\rangle$ & 0.161 &  0.160  \\
$\langle\phi^6\rangle$ & 0.130 &  0.133  \\ \hline
\multicolumn{1}{r}{R. S.} &\multicolumn{1}{r}{$\kappa =0.80$}
& \multicolumn{1}{c}{$s=10$} \\ \hline
$\langle A \rangle$                          & 1.50  & 1.50 \\
$\langle A^2 \rangle - \langle A \rangle ^2$ & 1.49 &  0.86 \\
$\langle S_{EC}\rangle $                                   & 1.35 &  1.24 \\
\hline \end{tabular}
\end{table}

We
perform a blocking transformation starting with our
bare action (which consists of just two operators), determine $T$
by measuring the appropriate correlators and compute $\nu$.
In doing that one needs the field anomalous dimension $\eta$.
$\eta$ is determined by demanding that at the fixed point the
renormalized and bare actions coincide.
It is of course
unlikely that the fixed point of this renormalization
group transformation lies just in the plane spanned by the two
operators of the bare action. However we can assess whether we are
close to it by adjusting $\eta$ and seeing how good the
agreement between the renormalized and the bare observables is.
Table 2 illustrates this point in a
$32^2 \to 16^2$ blocking. Similar results are
obtained in $64^2 \to 32^2$. For
comparison
some results for $\lambda \phi^4_3$ around the Wilson fixed point are also
shown. We conclude that we are still
at some distance from the fixed point and that a second renormalization
is probably necessary. (If we are close enough to the critical surface
the flow should drive us closer to the fixed point.)

\section{THE RESULTS}

It is a very pleasant surprise that the results turn out to be
extremely sensitive to the value of $\eta$ used to match the
`big' and `small' lattices. Our results from running on
a variety of values of $\kappa$ in the vicinity of the
crumpling transition with large statistics, both in the crumpled
and the smooth phases and on a variety of systems lead us
to conclude that $\eta=-1.05 \pm 0.05$. from the first blocking
transformation.

Our results from the second blocking are still very tentative and we
have only results for the $32^2\to 16^2 \to 8^2$ blocking, the
final lattice being undoubtedly too small. Even so, we get a value
$\eta\simeq
-1.3$, not too far from the one obtained in the first blocking. From
these two results we conclude that $-1.0\ge \eta\ge -1.3$  and
$3.1\le d_H \le 4.0$.

As for $\nu$, Table 3 shows the result
of the blocking $32^2\to 16^2\to 8^2$.
The second blocking yields very stable results. Should we
use these
values we would get $\nu=0.82$, but we shall not even quote
error bars here as these results are still very preliminary. Yet, they
show considerable promise, specially because the computational effort
required is relatively small. To mention some figures, to get the
above numbers requires about 1.5 hours on a Cray YMP or about
5 hours on a SGI Power Challenge L.

\begin{table}
\setlength{\tabcolsep}{1.pc}
\newlength{\digitwidth} \settowidth{\digitwidth}{\rm 0}
\catcode`?=\active \def?{\kern\digitwidth}
\caption{}
\label{tab:oper}
\begin{tabular}{rr} \hline
\multicolumn{1}{r}{\# of operators}
&\multicolumn{1}{c}{Largest eigenvalue} \\ \hline
\multicolumn{2}{c}{1st blocking} \\ \hline
 1 &  1.56  \\
 3 &  2.09  \\
 5 &  2.11  \\ \hline
\multicolumn{2}{c}{2nd blocking} \\ \hline
 1 & 1.99 \\
 3 & 2.22 \\
 5 & 2.23 \\ \hline
\end{tabular}
\end{table}

\end{document}